\documentclass{article}
\usepackage[T1]{fontenc} 
\usepackage[utf8]{inputenc} 
\usepackage{ismir,amsmath,cite,url}
\usepackage{graphicx}
\usepackage{color}
\usepackage{booktabs}
\usepackage{multirow}
 \usepackage{hhline}
\usepackage{xcolor}
\usepackage{adjustbox}
\usepackage{rotating}
\usepackage{tablefootnote}
\usepackage{threeparttable}
\usepackage{enumitem}
\usepackage{comment}
\usepackage{colortbl}
\newcommand{\yhyang}[1]{{\color{black}#1}} 
\newcommand{\sh}[1]{{\color{black}#1}} 
\newcommand{\jn}[1]{{\color{black}#1}} 
\newcommand{\anna}[1]{{\color{black}#1}} 

\title{EMOPIA: A Multi-modal Pop Piano Dataset for Emotion Recognition and Emotion-based Music Generation}

\multauthor
{Hsiao-Tzu Hung$^{1,2}$ \hspace{1cm} Joann Ching$^1$ \hspace{1cm} Seungheon Doh$^3$} { \bfseries{Nabin Kim$^4$ \hspace{1cm} Juhan Nam$^3$ \hspace{1cm} Yi-Hsuan Yang$^1$}\\
$^1$ Academia Sinica, Taiwan\\
$^2$ Department of CSIE, National Taiwan University\\
$^3$ Graduate School of Culture Technology, KAIST, South Korea\\
$^4$ Georgia Institute of Technology, United States\\{\tt\small r08922a20@csie.ntu.edu.tw, joann8512@citi.sinica.edu, seungheondoh@kaist.ac.kr}}

\def\authorname{H.T.Hung and J. Ching, and S.H Doh}

\usepackage[bookmarks=false,pdfauthor={\authorname},pdfsubject={\papersubject},hidelinks]{hyperref}

\begin{document}

\maketitle
\begin{abstract}
While there are many music datasets with emotion labels in the literature, they cannot be used for research on symbolic-domain music analysis or generation, as there are usually audio files only. 
In this paper, we present the EMOPIA (pronounced `yee-m\`{o}-pi-uh') dataset, a shared multi-modal (audio and MIDI) database focusing on perceived emotion in pop piano music, to facilitate research on various tasks related to music emotion. The dataset contains 1,087 music clips from 387 songs and clip-level emotion labels annotated by four dedicated annotators. Since the clips are not restricted to one clip per song, they can also be used for song-level analysis. 
We present the procedure for building the dataset, covering the song list curation, clip selection, and emotion annotation processes. Moreover, we prototype use cases on clip-level music emotion classification and emotion-based symbolic music generation by training and evaluating corresponding models using the dataset. The result demonstrates the potential of EMOPIA for being used in future exploration on piano emotion-related MIR tasks. 

\end{abstract}


\section{Introduction}\label{sec:introduction}

The affective aspect of music has been a major subject of research in the field of music information retrieval (MIR), not only for music analysis and labeling \cite{overviewEmotion,MER,GomezCanon2020ISMIR,GomezCanon21ISMIR,Bernardinis2020,panda2013multi,Panda2020,chowdhury19ismir,cancino2020express}, 
but also for music generation or editing \cite{ferreira,music_sketchnet,makris2021generating,sentimozart,Grimaud2020}. Accordingly, there have been quite a few public music datasets with emotion, 
as listed in Table \ref{table:emotion_dataset}.
These datasets are different in many ways, including the musical genres considered, data modality and data size, and the way emotion is described.

With the growing interest in symbolic-domain music analysis and generation in recent years of ISMIR \cite{chen20ismir,fadernets,pop909,dong2020muspy,mcLeod20ismir}, 
it is desirable to have an emotion-labeled symbolic music dataset to add emotion-related elements to such research.
However, among the datasets listed in Table \ref{table:emotion_dataset}, only two provide MIDI data, and they are both small in size.
Moreover, the majority of the audio-only datasets contain songs of multiple genres, making it hard to apply automatic music transcription algorithms, which currently work better for piano-only music \cite{o&f,TTtranscription,hawthorne2018enabling,cheuk2021reconvat}, to get MIDI-like data from the audio recordings.

\begin{table*}[h]
\centering
\begin{tabular}{l|llrl}
\toprule
\textbf{Name} & \textbf{Label type} & \textbf{Genre (or data source)} 
& \textbf{Size} & \textbf{Modality}  \\ 
\midrule
Jamendo Moods \cite{moodsThemes}    
& adjectives                                 & multiple genres                                                 & 18,486                                                                  & Audio                        \\
DEAM \cite{deam}                             & VA values                               & (from FMA \cite{fma_dataset}, Jamendo, MedleyDB\cite{medleydb})   
& 1,802                                                                  & Audio                        \\
EMO-Soundscapes \cite{emoSoundscape}                   & VA values                               & (from FMA)                                                    & 1,213                                                                  & Audio                \\
CCMED-WCMED \cite{ccmed} ~~                      & VA values                               & classical (both Western \& Chinese)           & 800                                                                   & Audio                        \\
emoMusic \cite{emoMusic}                          & VA values                           & pop, rock, classical, electronic                              & 744                                                                   & Audio                        \\
EMusic \cite{eMusic}                            & VA values                               & experimental, 8 others                                        & 140                                                                   & Audio                        \\
MOODetector \cite{panda2013multi}
& adjectives                             & multiple genres (AllMusic)                                         & 193
& Audio$+$MIDI   \\

VGMIDI  \cite{ferreira}                          & valence                                 & video game                                & 95 & MIDI                                  \\ 
\midrule
EMOPIA (\textbf{ours})             & Russell’s 4Q                    & pop (piano covers)                          & 1,078                        & Audio$+$MIDI  \\
\bottomrule
\end{tabular}
\caption{Comparison of some existing public emotion-labeled music datasets and the proposed EMOPIA dataset.}
\label{table:emotion_dataset}
\end{table*}

To address this need, we propose a new emotion-labeled dataset comprising of three main nice properties:
\vspace{1mm}
\begin{itemize}[leftmargin=*,itemsep=0pt,topsep=2pt]
\item \textbf{Single-instrument.} We collect audio recordings of piano covers and creations from YouTube, with fair to high audio and musical quality, and a diverse set of playing styles and perceived emotions. 
Focusing on only piano music allows for the use of piano transcription algorithms \cite{o&f,TTtranscription}, and facilitates disentanglement of musical composition from variations in timbre, arrangement, and other confounds seen in multi-instrument music.
\item \textbf{Multi-modal.} Both the audio and MIDI versions of the music pieces can be found from the Internet (see Section \ref{subsec:dataset_share} for details). The MIDI files are automatically transcribed from the audio by a state-of-the-art model \cite{TTtranscription}.
\item \textbf{Clip-level annotation.} The audio files downloaded from YouTube are full songs. As different parts of a song may convey different emotions, the first four authors of the paper manually and carefully pick emotion-consistent short clips from each song and label the emotion of these clips using \anna{a four-class taxonomy derived from the Russell's valence-arousal model} \cite{russell}.
This leads to clip-level emotion annotations for in total 1,087 clips from 387 songs (i.e., 2.78 clips per song on average), with the number of clips per emotion class fairly balanced.

\end{itemize}



Given these properties, EMOPIA has versatile use cases in MIR research.
For music labeling, EMOPIA can be used for clip-level music emotion recognition or music emotion variation detection \cite{MER}, in both the audio and symbolic domains. 
For music generation, EMOPIA can be used for emotion-conditioned piano music generation or style transfer, to create emotion-controllable new compositions, or variations of existing pieces, again in both domains. 

We present details of the dataset and the way we compile it in Section \ref{sec:dataset_creation}, and report a computational analysis of the dataset in Section \ref{sec:dataset_analysis}.
Moreover, to demonstrate the potential of the new dataset, we use it to train and evaluate a few clip-level emotion classification models using both audio and MIDI data 
in Section \ref{sec:classification}, and emotion-conditioned symbolic music generation models in Section \ref{sec:generation}. The latter involves the use of a recurrent neural network (RNN) model 
proposed by Ferreira \emph{et al.} \cite{ferreira}, and our own modification of a Transformer-based model \cite{transformer,huang2020pop,hsiao2021compound} that takes emotion as a conditioning signal for generation.

\yhyang{We release the EMOPIA dataset at a Zenodo repo.\footnote{\url{https://zenodo.org/record/5090631}}
Source code implementing the 
generation and classification  models can be found on GitHub.\footnote{\url{https://github.com/annahung31/EMOPIA}}\footnote{\url{https://github.com/SeungHeonDoh/EMOPIA_cls}}
Examples of generated pieces  can be found at a demo webpage.\footnote{\url{https://annahung31.github.io/EMOPIA/}}}



\section{Related Work}\label{sec:related_work}

\textbf{Emotion Recognition in Symbolic Music}.
Symbolic music representations describe music with the note, key, tempo, structure, chords, instruments. To understand the relationship between music and emotion, various researchers have investigated machine learning approaches with handcrafted features. Grekow \emph{et al.} \cite{grekow2009detecting} extract in total 63 harmony, rhythmic, dynamic features from 83 classic music MIDI files. Lin \emph{et al.} \cite{lin2013exploration} compare audio, lyric, and MIDI features for music emotion recognition, finding that MIDI features lead to higher performance in valence dimension. Panda \emph{et al.} \cite{panda2013multi} also proposed multi-model approaches, combining audio
and MIDI features for emotion recognition using a small dataset of 193 songs. 

\vspace{3mm}
\noindent\textbf{Emotion-conditioned Symbolic Music Generation}. 
Only few work has started to address this task recently.
Ferreira \emph{et al.} \cite{ferreira} compile a small dataset of video game MIDI tracks with manual annotations of valence values, named VGMIDI (cf. Table \ref{table:emotion_dataset}), and use it to train a 
long short term memory network (LSTM) in tandem with a genetic algorithm (GA) based elite selection mechanism to generate positive or negative music. 
Makris \emph{et al.} \cite{makris2021generating} approach the same task by using designated chord progression sequence in a  sequence-to-sequence architecture trained with the VGMIDI dataset.
Zhao \emph{et al.} \cite{ITNEC2019lstm} use LSTM to generate music with four different emotions.
Madhok \emph{et al.} \cite{sentimozart} use human facial expressions as the condition to generate music. 
More recently, Tan \& Herremans demonstrate that their FaderNets \cite{fadernets} can achieve arousal-conditioned symbolic music style transfer with a semi-supervised clustering method that learns the relation between high-level features and low-level representation.
Their model modifies the emotion (specifically, only the arousal) of a music piece, instead of generating new pieces from scratch.



\begin{figure*}
\centering
\includegraphics[width=\textwidth]{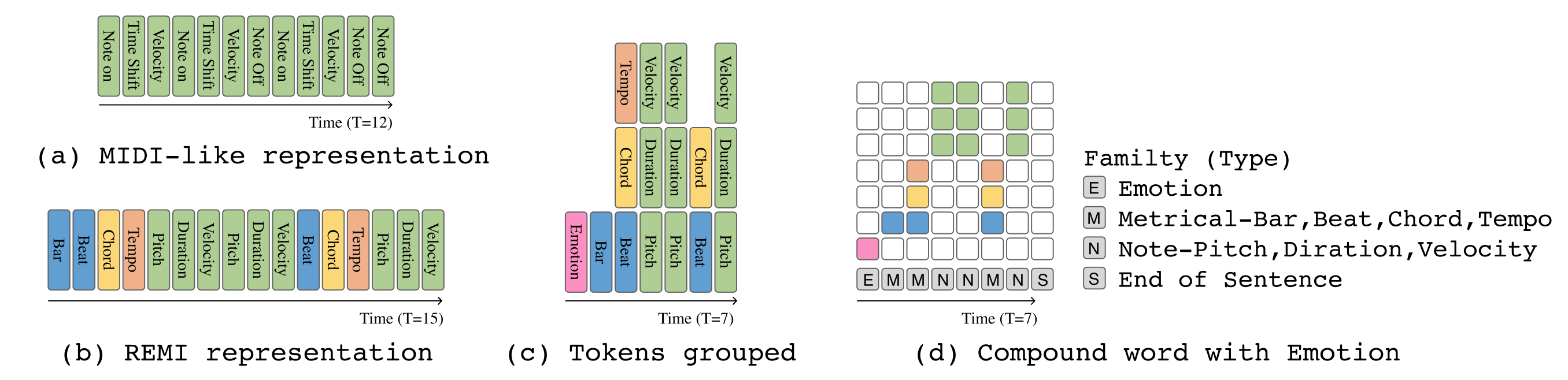}
\caption{Illustration of different token-based representation for symbolic music: (a) MIDI-like \cite{oore2018time}, (b) REMI \cite{huang2020pop}, and (d) CP \cite{hsiao2021compound} plus emotion token. Sub-figure (c) is an intermediate representation of the CP one.}
\label{fig:token}
\end{figure*}

\section{The EMOPIA Dataset}\label{sec:dataset_creation}


\subsection{Song Selection and Segmentation}

EMOPIA is a collection of 387 
piano solo performances of popular music 
segmented manually into 1,087 clips for emotion annotation.
Two authors of the paper curated the song list of the piano performances by scanning through playlists on Spotify for its consistently high quality, then downloading the recordings from YouTube.
A song is included when it \jn{is played by a professional conveying a clear emotion, and the recording has not been heavily engineered during post-production.}
The genres of songs include Japanese anime, Korean and Western pop song covers, movie soundtracks, and personal compositions.


In an effort to extend the usefulness of the dataset for future research, at the best of our ability, the songs are intentionally segmented (with the help of the Sonic Visualizer \cite{sonicvisualizer}) only at cadential arrivals to make it an \emph{emotionally-consistent clip} and a valid \emph{musical phrase} at the same time. Accordingly, EMOPIA contains information for full songs, extracted phrases, and emotion labels.

\subsection{Emotion Annotation}
\label{annotate}

\anna{Different emotion taxonomies have been adopted in the literature for emotion annotation, with no standard so far \cite{MER}. 
For EMOPIA, we consider 
a simple four-class taxonomy corresponding to the four quadrants of the Russell's famous Circumplex model of affect \cite{russell}, which conceptualizes emotions in a two-dimensional space defined by valence and arousal. 
The four classes are:  HVHA (high valence high  arousal); HVLA (high valence low arousal); LVHA (low valence high arousal); and LVLA (low valence low arousal). We  refer to this taxonomy as Russell's 4Q.} 

As pointed out in the literature \cite{cultureEmotion,MER,GomezCanon2020ISMIR}, various factors affect the perceived emotion of music, including cultural background, musical training, gender, etc. Consensus on the perception of emotion is challenging accordingly. Therefore, the annotations were made only among \jn{the first four authors}, all coming from similar cultural backgrounds and collaborating closely during the annotation campaign, to \jn{ensure mutual standards for high or low valence/arousal}.
As it is time-consuming and laborious to choose clips from a song and label the emotion, each song was taken care of by only one annotator. Yet, cross validation of the annotations among the annotators was made several times during the annotation campaign (which spans 2.5 months), to ensure all annotators work on the same standard.

Table \ref{tab:clip_num} shows the number of clips and the average length (in seconds) for each class. The clips 
amount to approximately 11 hours’ worth of data. 


\begin{table}
\centering
\begin{tabular}{l|rr} 
\toprule
 \multicolumn{1}{l|}{\textbf{Quadrant}}   & \multicolumn{1}{l}{\textbf{\# clips}} & \multicolumn{1}{l}{\textbf{Avg. length} (in sec~/~\#tokens)}  \\ 
\midrule
Q1 & 250                       & 31.9~/~1,065 \\ 
Q2 & 265                       & 35.6~/~1,368\\ 
Q3 & 253                       & 40.6~/ ~~~771\\ 
Q4 & 310                       & 38.2~/ ~~~729\\
\bottomrule

\end{tabular}
\caption{\anna{The number of clips and their average length in seconds, or in the number of REMI tokens,} for each quadrant of the \anna{Russell's model} in the EMOPIA dataset.}
\label{tab:clip_num}
\end{table}

\subsection{Transcription}

We transcribe the selected clips automatically with the help of the high-resolution piano transcription model proposed by Kong \emph{et al.} \cite{TTtranscription}, which is open source and represents the state-of-the-art for this task. We have manually checked the transcription result for a random set of clips and find the accuracy in note pitch, velocity, and duration satisfactory. 
The transcription  might be fragmented and undesirable for cases such as when the audio recording is engineered to have unnatural ambient effects; 
we drop such songs from our collection.
The model also transcribes pedal
information, which we include to EMOPIA but do not use in our experiments.

\begin{figure*}
\centering
\includegraphics[width=.9\textwidth]{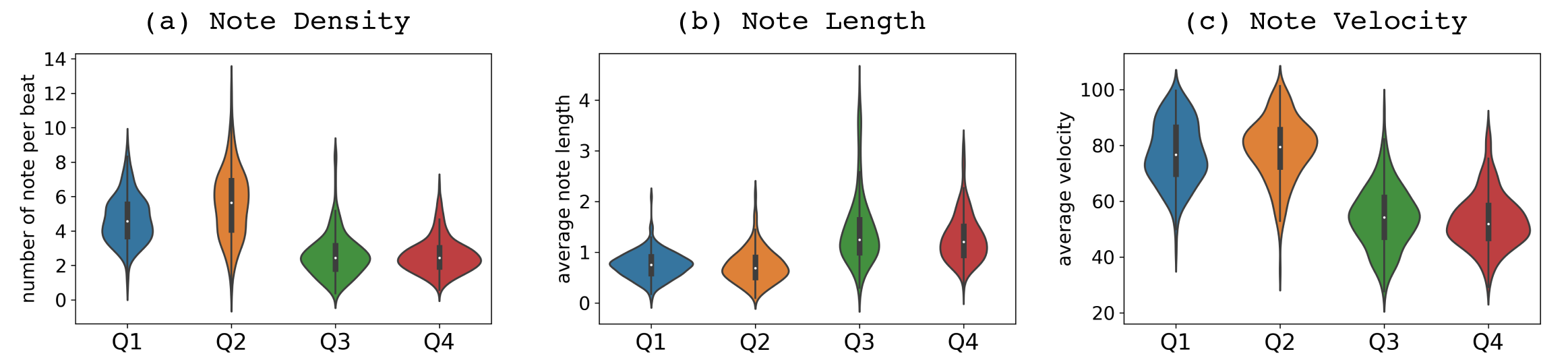}
\caption{Violin plots of the distribution in  (a) note density, (b) length, and (c)  velocity for clips from different  classes.}
\label{fig:stat}
\end{figure*}

\begin{figure}[!t]
\centering
\includegraphics[width=\linewidth]{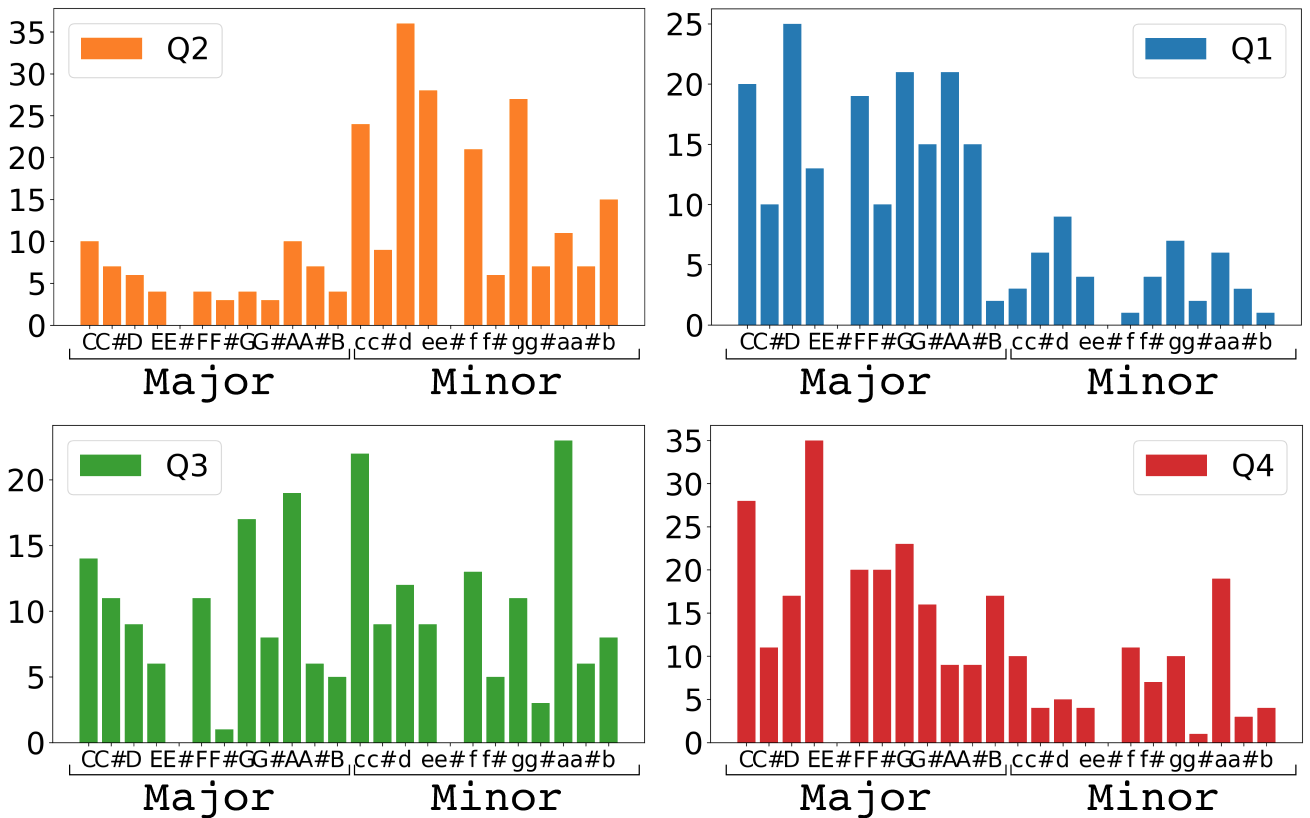}
\caption{Histogram of the keys (left~/~right: major~/~minor keys) for  clips from different emotion classes.}
\label{fig:key}
\end{figure}

\subsection{Pre-processing and Encoding}
\label{encoding}



For building machine learning models that deal with symbolic data, we need a data representation that can be used as input to the models. For example, MusicVAE  \cite{roberts2018hierarchical} adopts the \emph{event-based} representation that encodes a symbolic music piece as a sequence of ``event tokens'' such as note-on and note-off, while MuseGAN \cite{dong2018musegan} employs a timestep-based, \emph{piano roll}-like representation.
Since there is no standard on the symbolic representation thus far, we adopt the following event-based ones in our experiments. Specifically, we use MIDI-like and REMI in Section \ref{sec:classification} and CP in Section \ref{sec:generation}. See Figure \ref{fig:token} for illustrations.
\vspace{1mm}
\begin{itemize}[leftmargin=*,itemsep=0pt,topsep=2pt]

\item The \textbf{MIDI-like} representation  \cite{oore2018time} encodes information regarding a MIDI note with a ``note-on'' token, a ``note-off'', and a ``velocity'' token. Moreover, the ``time shift'' token is used to indicate the relative time gap (in ms) between two tokens.

\item \textbf{REMI} \cite{huang2020pop} considers a beat-based representation that instead uses ``bar'' and ``subbeat'' tokens to represent the time information. A ``bar'' token signifies the beginning of a new bar, and ``subbeat'' points to one of the constant number of subbeat divisions in a bar.
Additionally, REMI uses ``tempo'' tokens to control the pace of the music, and replaces ``note-off'' with ``note-duration''.
Table \ref{tab:clip_num} also shows the average number of REMI tokens for clips in each emotion class.

\item \textbf{CP} \cite{hsiao2021compound}. Both MIDI-like and REMI view events as individual tokens.
In CP, tokens belonging to the same \emph{family} are grouped into a \emph{super token} and placed on the same timestep (see Figure \ref{fig:token}(c)). CP considers by default three families: \emph{metrical}, \emph{note}, and \emph{end-of-sequence}. We additionally consider the ``emotion'' tokens and make it a new family, as depicted in Figure \ref{fig:token}(d). 
The \emph{prepending} approach is motivated by  CTRL \cite{keskarCTRL2019}, a state-of-the-art controllable text generation model in natural language processing (NLP) that uses global tokens to affect some overall properties of a sequence.

\end{itemize}


\subsection{Dataset Availability}
\label{subsec:dataset_share}
In EMOPIA, each sample is accompanied with its corresponding metadata, segmentation annotations, emotion annotation, and transcribed MIDI, \yhyang{which are all available in the Zenodo repository.}
\anna{Moreover, we have added the MIDI data to the MusPy library \cite{dong2020muspy} to facilitate its usage.}
Due to copyright issues, however, we can only share audio through YouTube links instead of sharing the audio files directl; the availability of the songs are subject to the copyright licenses in different countries and whether the owners will remove them.

\begin{table}[!t]
\begin{adjustbox}{width=\linewidth}
\begin{tabular}{l|cc|cc}
\toprule
\multirow{2}{*}{Label pair} & \multicolumn{2}{c|}{\yhyang{Different arousal}} & \multicolumn{2}{c}{\yhyang{Different valence}}    \\ 
    & Q1 vs. Q4  & Q2 vs. Q3          & Q1 vs. Q2             & Q3 vs. Q4  \\
\midrule
JS Div. & 0.236           & 0.250          & 0.434             & 0.280            \\ 
\bottomrule
\end{tabular}
\end{adjustbox}
\caption{\yhyang{Jensen-Shannon divergence between the key histograms of a few  emotion quadrant pairs. }}
\label{tab:key-JS}
\end{table}

\section{Dataset Analysis}\label{sec:dataset_analysis}
The emotion that listeners perceive is determined by a wide array of composed and performed features of music \cite{Juslin2011}. To observe the emotional correlate of the musical attributes in EMOPIA, we extract various MIDI-based features and examine the distributions over the four quadrants of emotion. We present here the most discriminative features among many choices we examined in our analysis. 
 

\vspace{3mm}
\noindent \textbf{Note Density, Length, and Velocity}. The arousal of music can be easily observed based on the frequency of note occurrences and their strength \cite{livingstone2010changing}. We measure them by note density, length and velocity.  The note density is defined as the number of notes per beat, and the note length is defined as the average note length in beat unit. The note velocity is obtained directly from MIDI. Figure \ref{fig:stat} shows three violin plots of the three features. The general trend shows that the high-arousal group (Q1, Q2) and low-arousal group (Q3, Q4) are distinguished well in all three plots. The results of note density and velocity are expected considering the nature of arousal. However, it is quite interesting that the note lengths are generally longer in the low-arousal group (Q3, Q4). Between the same-arousal quadrants, the differences are subtle. In note density, Q2 has more dynamics than Q1, whereas Q3 is not distinguishable from Q4. In note length, Q1 has slightly longer notes than those of Q2, whereas Q3 is again not distinguishable from Q4. In velocity, Q2 have louder notes than those of Q1, and Q3 has slightly louder notes than Q4.    


\vspace{3mm}
\noindent \textbf{Key Distribution}. 
The valence of music is often found to be related to the major-minor tonality \cite{Panda2020}.
For simplicity, we measure the tonality from the musical key. We extract the key information using the Krumhansl-Kessler algorithm \cite{krumhansl2001cognitive} in the \text{MIDI ToolBox} \cite{Eerola2004Toolbox}. Figure \ref{fig:key} shows the key distributions on 12 major-minor pitch classes for the four emotion quadrants. They assure that the major-minor tonality is an important clue in distinguishing valence. In the high valence group (Q1, Q4), the distribution is skewed to the left (major), while, in the low valence group (Q2, Q3), the trend is the opposite. We also measured the distance between the key distributions using the Jensen-Shannon divergence, which has the property of symmetry. Table \ref{tab:key-JS} summarizes the pair-wise distances, indicating that the valence difference group has a \sh{larger} difference than the arousal difference group.   







\section{Music Emotion Recognition}\label{sec:classification}
We report our baseline research on both symbolic- and audio-domain emotion recognition using EMOPIA, which defines the task as classifying a song into four categories. 
The clips in EMOPIA are divided into train-validation-test splits with the ratio of 7:2:1 in a stratified manner.

\begin{table}[t]
\centering
\begin{tabular}{l|ccc}
\toprule
\textbf{Model}  & \textbf{4Q} & \textbf{A} & \textbf{V} \\ \midrule
Logistic regression  & \sh{.581} & \sh{.849} & \sh{.651} \\ \midrule
LSTM-Attn~\cite{lin2017structured}$+$MIDI-like~\cite{oore2018time} & \sh{\textbf{.684}} & \sh{.882} & \sh{\textbf{.833}} \\ 
LSTM-Attn~\cite{lin2017structured}$+$REMI\cite{huang2020pop} & \sh{.615} & \sh{\textbf{.890}} & \sh{.746} 
 \\  \bottomrule
\end{tabular}
\caption{Symbolic-domain classification performance.}
\label{tab:classification-midi}
\end{table}

\begin{table}[t]
\centering
\begin{tabular}{l|ccc}
\toprule
\textbf{Model} 
& \textbf{4Q} & \textbf{A} & \textbf{V} \\ \midrule
Logistic regression  & \sh{.523} & \sh{\textbf{.919}} & \sh{.558} \\
\midrule
Short-chunk ResNet \cite{won2020evaluation} & \sh{\textbf{.677}} & \sh{.887} & \sh{\textbf{.704}} \\   \bottomrule
\end{tabular}
\caption{Audio-domain classification performance.}
\label{tab:classification-audio}
\end{table}


\vspace{3mm}\noindent\textbf{Symbolic-domain Classification}. 
We evaluate two methods: one is based on hand-crafted features with a simple classifier, and the other is on a deep neural network model. For the former, we use the analysis features in the previous section and a logistic regression classifier as a baseline. Specifically, we use average values of note density, note length, velocity, and represent the key as a one-hot vector. For the latter, we use two different symbolic note representation methods introduced in Section \ref{encoding}, 
MIDI-like \cite{oore2018time} and REMI \cite{huang2020pop}. 
For the learning model, we use the combination of bidirectional LSTM and a self-attention module, or \emph{LSTM-Attn} for short, proposed originally for sentiment classification in NLP  \cite{lin2017structured}. The LSTM extracts temporal information from the MIDI note events, while the self-attention module calculates different weight vectors over the LSTM hidden states with multi-head attentions. The weighted hidden states are finally used for classification.  


\vspace{3mm}\noindent\textbf{Audio-domain Classification}. We evaluate two audio-domain classification methods in a similar manner to the symbolic-domain ones. In the first method, we use an average of 20 dimensions of mel-frequency cepstral coefficient (MFCC) vectors and a logistic regression classifier. In the second method, we use the short-chunk ResNet following \cite{won2020evaluation}, which is composed of 7-layer CNNs with residual connections.  The output is summarized as a fixed 128-dimensional vector through max pooling, which is followed by two fully connected layers with the ReLU activation for classification. The input audio to the short-chunk ResNet is 3-second excerpts represented as a log-scaled mel-spectrogram with 128 bins with 1024-size FFT (Hanning window), and 512 size hop at 22,050 Hz sampling rate. 
We randomly sample three seconds of the audio chunk as an input size to the classifier. 

\vspace{3mm}\noindent\textbf{Evaluation}. We calculate 4-way classification accuracy over the four different emotion quadrants and 2-way classification accuracy over either arousal and valence. \sh{ Tables \ref{tab:classification-midi} and \ref{tab:classification-audio} show the results in the symbolic and audio domains, respectively. Except for arousal, we can see that the deep learning approaches generally outperform the logistic regression classifiers using hand-crafted features. In audio domain arousal classification, MFCC vectors averaging the entire song sequence showed better performance than the deep learning approach with 3-second input. It seems that wider input sequence has the strength in emotion recognition. In both domains, valence classification is a more difficult task compared to arousal classification. 
For valence classification, MIDI-domain classifiers yield better result than audio-domain classifiers (0.883 vs. 0.704).} 
\yhyang{Among the two token representations, MIDI-like seems to outperform REMI for valence classification.}

\begin{table*}[h]
\centering
\begin{tabular}{l|ccc|ccc|ccc}
\toprule
\multirow{2}{*}{\textbf{Model}} & \multicolumn{6}{c|}{\textbf{Objective metrics}} & \multicolumn{3}{c}{\textbf{Subjective metrics}} \\
\cline{2-7} \cline{8-10}
&   {PR}   & {NPC}   & {POLY}  & {4Q} & {A} & {V} &  {Humanness}  & {Richness}  & {Overall}\\ \midrule
EMOPIA (i.e., real data)  & 51.0   & 8.48  & 5.90 & --- & --- & --- & ---  & ---  & ---\\ 
\midrule
LSTM$+$GA \cite{ferreira} & 59.1  & 9.27 & 3.39 & \jn{.238} & \jn{.500} & \jn{.498} & 2.59\small{$\pm$1.16}  & 2.74\small{$\pm$1.12}  & 2.60\small{$\pm$1.07}\\
CP Transformer \cite{hsiao2021compound} & 53.4    & 9.20  & 3.48  & \textbf{\anna{.418}} & \textbf{\anna{.690}} & \anna{.583} & 2.61\small{$\pm$1.03}  & 2.81\small{$\pm$1.03}  & 2.78\small{$\pm$1.03}\\
CP Transformer w/ pre-training & \textbf{49.6} & \textbf{8.54}  & \textbf{4.40} & \anna{.403} & \anna{.643} & \textbf{\anna{.590}}  & \textbf{3.31\small{$\pm$1.18}}  & \textbf{3.22\small{$\pm$1.23}}  & \textbf{3.26\small{$\pm$1.15}}\\ 
\bottomrule
\end{tabular}
\caption{Performance comparison of the evaluated models for emotion-conditioned symbolic music generation in 
\emph{surface-level objective metrics} (\underline{P}itch \underline{R}ange, \underline{N}umber of \underline{P}itch \underline{C}lasses used, and \underline{POLY}phony; the closer to that of the real data the better), 
\emph{emotion-related objective metrics} (\underline{4Q} classification, \underline{A}rousal classification, \underline{V}alence classification; the higher the better), 
and \emph{subjective} metrics  (all in 1--5; the higher the better); bold font highlights the best result per metric.} 
\label{tab:obj}
\end{table*}

\section{Emotion-conditioned Generation}\label{sec:generation}

We build the Transformer and LSTM models for emotion-conditioned symbolic music generation using EMPOIA. 
For the former, 
we adopt the Compound Word Transformer \cite{hsiao2021compound}, the state-of-the-art in unconditioned symbolic music generation. We employ the CP$+$emotion representation presented in Section \ref{encoding} as the data representation. 

For the LSTM model, we consider the approach proposed by Ferreira \emph{et al.} \cite{ferreira}, which represents the state-of-the-art in emotion-conditioned music generation. 
Our implementation follows that described in \cite{ferreira}, \yhyang{with the following differences: 1) train on EMOPIA rather than VGMIDI; 2) use 512 neurons instead of 4,096 due to our limited computational resource; 3) use the same linear logistic regression layer for classification but we classify four classes instead of two.}


As the size of EMOPIA might not be big enough, we 
use additionally the \emph{AILabs1k7} dataset compiled by Hsiao \emph{et al.} \cite{hsiao2021compound} to pre-train the Transformer. AILabs1k7 contains 1,748 samples and is also pop piano music, but it does not contain emotion labels.
Most of the clips in AILabs1k7 are longer than EMOPIA, so to keep the consistency of the input sequence length, the length of the token sequence is set to be 1,024. 
We pre-train the 
Transformer with 1e--4 learning rate on AILabs1k7, take the checkpoint with negative log-likelihood loss 0.30, and then fine-tune it on EMOPIA with 1e--5 learning rate. During pre-training, the emotion token is always set to be ``ignore,'' while in fine-tuning it is set to the emotion of that sample.

\vspace{3mm}\noindent\textbf{Evaluation}. We use the following three sets of metrics. 
\vspace{1mm}
\begin{itemize}[leftmargin=*,itemsep=0pt,topsep=2pt]

\item \textbf{Surface-level objective metrics.} We use the following three metrics proposed by Dong \emph{et al.} \cite{dong2018musegan} to evaluate whether the generated samples fit the training data: {pitch range} (PR),  {number of unique pitch classes used} (NPC), and {number of notes being played concurrently} (POLY). We use MusPy \cite{dong2020muspy} to compute these metrics.

\item \textbf{Emotion-related objective metrics.} 
\yhyang{Since both REMI and CP adopt a beat-based representation of music, we employ the LSTM-Attn$+$REMI emotion  classifier} (cf. Section \ref{sec:classification})  here to quantify how well the generation result is influenced by the emotion condition. We first use the generative model to generate 100 samples per class, and use the assigned label as the target class of the sample. The trained classifier is then used to make prediction on the generated samples. Similar to the classification task, apart from 4Q classification, we also conduct 2-way classification of both Arousal and Valence aspect. 

\item \textbf{Subjective metrics.}  As the classifiers are not 100\% accurate, we also resort to a user survey to evaluate the 
emotion-controllability of the  models. Specifically, we deploy an online survey to collect responses to the music generated by different models. A subject has to listen to 12 random-generated samples, one for each of the three models and each of the four emotion classes, and rate them on a five-point Likert scale with respect to 1)~Valence: is the audio negative or positive; 2)~Arousal: is low or high in arousal; 3)~Humanness: how well it sounds like a piece played by human; 4)~Richness: is the content interesting; and, 5)~Overall musical quality. 
In total 25 subjects participated in the survey. 
\end{itemize}
\vspace{2mm}
Table \ref{tab:obj} tabulates some of the results.
We see that the CP Transformer with (`w/') pre-training performs the best in \yhyang{most of the}  
objective metrics and the three subjective metrics listed here.
\yhyang{Nevertheless, the scores of the CP Transformer with pre-training in the three emotion-related objective metrics are much lower than that reported in Table \ref{tab:classification-midi}, suggesting that either the generated pieces are not emotion-laden, or the generated pieces are too dissimilar to the real pieces to the classifier.}




\begin{figure}[!t]
\centering
\includegraphics[width=0.8\linewidth]{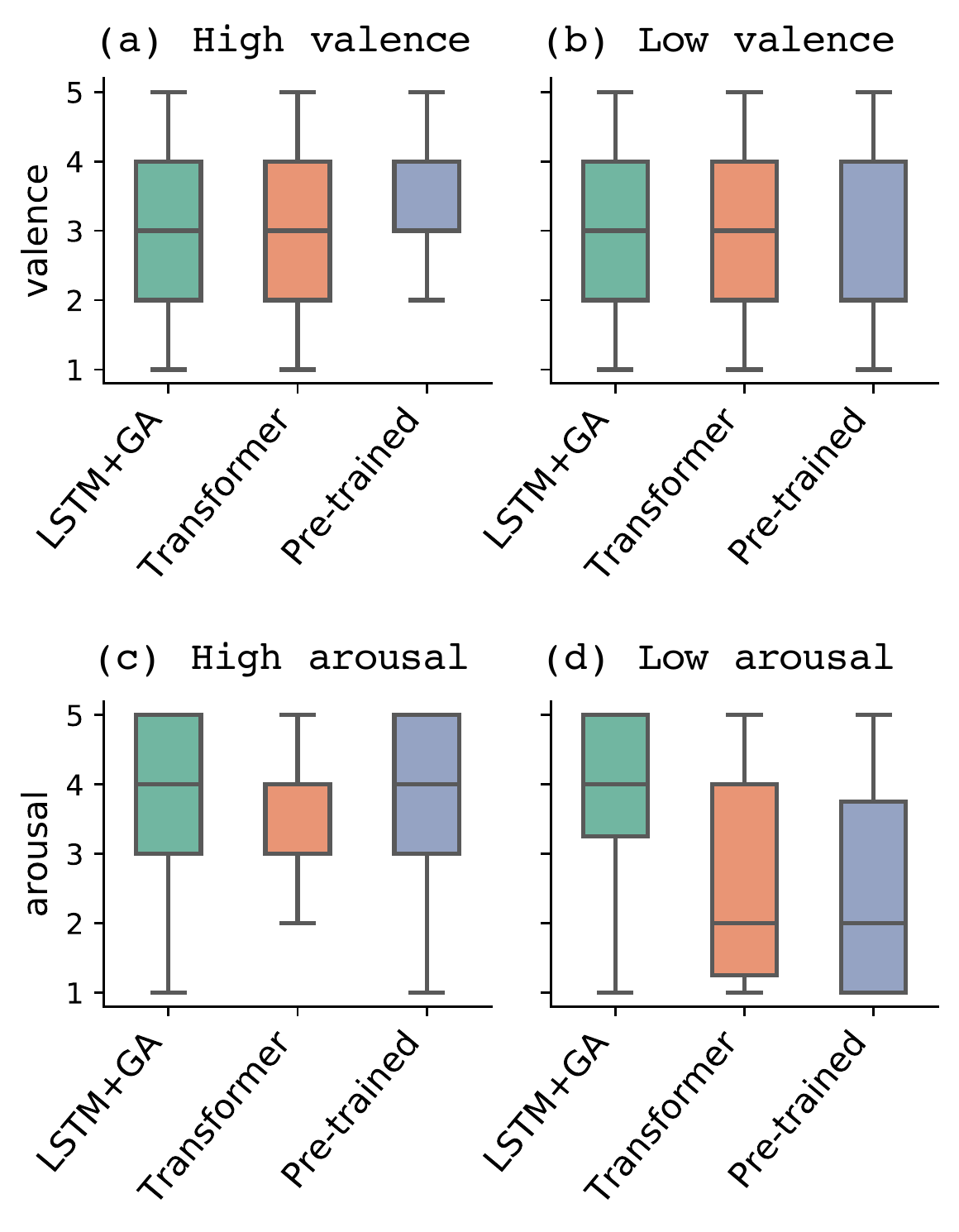}
\caption{The subjective emotional scores (in 1--5) for the generative results when the target emotion is (a) high valence, (b) low valence, (c) high arousal,  (d) low arousal. } 
\label{fig:userstudy}
\end{figure}



Figure \ref{fig:userstudy} shows the human assessment of the emotion-controllability of the models. 
\yhyang{To our surprise, while the CP Transformer with pre-training does not score high in the emotion related objective metrics, the subjective test shows that it can actually control the emotion of the generated pieces to a certain extent, better than the two competing models.}
In particular, the valence of the samples generated by the Transformer with pre-training has a median rating of 4 when the goal is to generate positive-valence music (i.e., Figure \ref{fig:userstudy}(a)), while the scores of the other two models are around 3. 
Moreover, the arousal of the samples generated by the Transformer with pre-training has a median rating of 4 when the goal is to generate high-arousal music (Figure \ref{fig:userstudy}(c)), which is higher than that of the non pre-trained Transformer.
\yhyang{This suggests that the LSTM-Attn classifier employed for computing the emotion-related objective metrics may not be reliable enough to predict the emotion perceived by human, and that the Transformer with pre-training is actually effective in controlling the emotion of the music it generates to certain extent.}
But, the model seems not good enough for the cases of generating low-valence (i.e., negative) music, as shown in Figure \ref{fig:userstudy}(b). 

\vspace{-2mm}
\section{Conclusion}

In this paper, we have proposed a new public dataset EMOPIA, a medium-scale emotion-labeled pop piano dataset. 
It is a multi-modal dataset that contains both the audio files and MIDI transcriptions of piano-only music, along with clip-level emotion annotations in four classes. 
We have also presented prototypes of models for clip-level music emotion classification and emotion-based symbolic music generation trained on this dataset, using a number of state-of-the-art models in respective tasks. The result shows that we are able to achieve high accuracy in both four-quadrant and valence-wise emotion classification, and that our Transformer-based  model is capable of generating music with a given target emotion to a certain degree.

In the future, in-depth importance analysis can be conducted to figure out features that are important for emotion classification, and to seek ways to incorporate those features to the generation model. Many ideas can also be tried to further improve the performance of emotion conditioning, e.g., the \anna{Transformer-GAN} approach \cite{muhamed2021symbolic}. 


We share not only the dataset itself but the code covering all our implemented models in a GitHub repo.
We hope that researchers will find this contribution useful in future emotion-related MIR tasks.




\bibliography{ISMIRtemplate}

%
%
%
%
%

\end{document}